# Imbibition: An Example of Nonconserved Cellular Automaton

P. B. Sunil Kumar* and Debnarayan Jana†
*Raman Research Institute, Bangalore 560 080, India*

We report an imbibition experiment in 2D random porous media in which height - height correlation function grows with a nonuniversal exponent. We find the exponent to depend on evaporation. A cellular automaton model for imbibition based on first principles is presented. A numerical study of the model gives results which are consistent with the experiment. The interface is shown to exhibit self-organised criticality (SOC).

## I. INTRODUCTION

Interface growth phenomena has been studied in a variety of systems [1–3]. Recently viscous imbibition experiments with water displacing air in a Hele-Shaw cell filled with glass beads opened up a new area to study interface growth [4]. Of late, another kind of imbibition experiments was studied to understand interface growth [5–7]. In these experiments a paper is fixed with the bottom end dipped into a suspension. The suspension is imbibed into the pores of the paper by capillary action. The fluid rises through the pores carrying the suspended particles with it. Also, as the fluid rises, it continuously gets evaporated from the surface of the paper. An interface is formed by the wet front which rises steadily. There is some amount of randomness present in the paper due to the blocked pores. This randomness and evaporation of the fluid try to pin this wetting front. The motion of the wet front is impeded not only by the evaporation rate but also by the concentration of the suspension, the ratio of the size of the suspended particles to the pore size and viscosity of the fluid. The front stops moving when the fluid at the boundary has completely evaporated. These experiments provide a simple laboratory system which mimics diffusion in random porous media. These are of importance in chromatography. They could also provide good "table top" systems to study pattern formation [8,9].

There are some qualitative differences between this interface formation in imbibition and other growth models. The roughness of the interface in imbibition is solely due to the disorder in the paper. If there is no blocking of the pores the interface will be smooth at any rate of evaporation. However, the disorder experienced by the fluid is indirectly affected by the evaporation. Remember there could be many paths to reach a particular cell. When the evaporation is small the fluid can take even longer paths to reach a cell. This limits the effectiveness of the randomness of the medium in pinning the interface. For the same amount of randomness the number of paths will be lesser when the evaporation is higher, because higher evaporation eliminates longer paths. Thus randomness asserts more when evaporation is higher. It should be mentioned that this imbibition case differs from the deposition models in three ways (i) there is a time scale set by the evaporation (ii) this evaporation is from the bulk not from the interface alone. Thus evaporation couples the interface to the bulk and (iii) the randomness in imbibition is quenched and not fluctuating in time.

It is known that the wet front describes a self-affine fractal roughness [1]. We fix the geometry of the system under consideration as follows. Assume that the interface to be oriented in the $x$ direction. $h(x)$ is the height variable along $x$. The roughness of the interface can then be characterised by the exponent $\alpha$ defined as

$$W(l) \sim l^\alpha \qquad (1)$$

where, $W(l)$ is the correlation of the fluctuations in the height at two points $x$ and $x + l$ after the front stops moving and is given by

$$W(l) = \left\langle (h(x) - h(x+l))^2 \right\rangle_x^{1/2} \qquad (2)$$

Till now most of the understanding of the interface growth phenomeona in imbibition comes through computer simulations. So far no simulations in imbibition have been done taking care of evaporation, size of the particles and concentration of the suspension explicitly. In the treatment by Amaral *et al* [6] evaporation was incorporated phenomenologicaly by a steady increase ($\Delta p$) at each time step in the probability $p$ of blocking the pores. This increase drives the system into a percolation threshold $p_c$ wherein they obtained a connected cluster of blocked pores. This treatment assumes a linear increase of $p$ with time. In an alternative description one can associate a dimensionless driving force $f = (F - F_c)/F_c$ with the interface, with $F$ decreasing to $F_c$ being equivalent to $p \to p_c$. That is the change of $p$ in this model is in effect a linear decrease of this driving force $f$ with time, driving it to $F_c$. By incorporating evaporation explicitly we show that a nontrivial nonlinear dependence of driving force will be more appropriate.

It is believed that in general interface growth can be described by a nonlinear growth equation [10,11]

$$\partial_t h(x) = F + \nu \nabla^2 h(x) + \lambda(\nabla h(x))^2 + \zeta(x,h) \qquad (3)$$

where, $\zeta(x,h)$ is quenched noise and $F$ is a driving force.

The above equation describes two universality classes depending on the value of $\lambda$ as $F \to F_c$. In the case of Directed Percolation Depinning (DPD) $\lambda \to \infty$ while in quenched Edward - Wilkinson (QEW) universality class $\lambda \to 0$ as $F \to F_c$. The model treated by Amaral et. al. [6] falls into the DPD universality class. However, we find imbibition to be different from these two classes by the presence of a driving force which tunes itself close to $F_c$ leading to self-organised crticality (SOC).

In general SOC is characterised by the emergence of scale invariance in an extended nonequilibrium system [12]. Even though there is a substantial amount of work in this field till now there is no general aggrement on the origin and characterisation of SOC in different systems. In fact conservation laws were believed to be necessary for the appearance of SOC [13]. However there are also models which have nonconservation but still shows SOC [14,15]. The model presented here belongs to this class of nonconserved systems exhibiting SOC.

In this paper we present a detailed analysis of the results communicated through a short paper [7]. The paper is organised as follows. In section 2 we describe an imbibition experiment at two different experimental conditions. In section 3 we present two microscopic models to study imbibition. A numerical study of the models and comparison of the results with that of experiments is given in section 4. We point out an interesting connection beteween imbibition and other nonconserved models and finally in section 5 we give our conclusions.

## II. EXPERIMENT

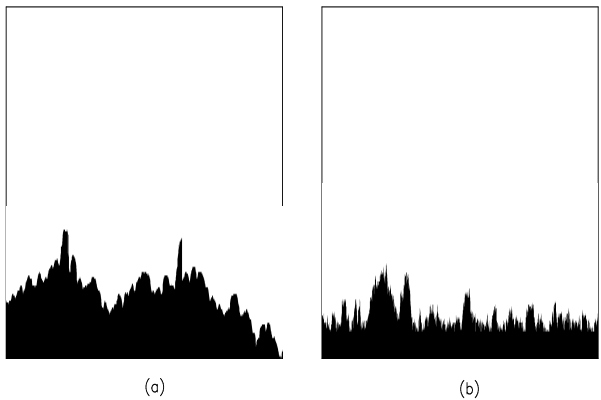

FIG. 1. (a) A digitized image of the interface from experiment. (b) A typical result of the simulation obtained using model I

The experiments are carried out using Whatman No:1 filter paper as the porous medium and ink as the suspension. The evaporation rate was varied by changing the room humidity and temperature. The ink rises through the paper and stops at a particular height. The darkening of the interface is indicative of this stoppage of growth. This darkening is due to the following reason. As the ink rises the fluid evaporates. As a result the suspended particles carried by the fluid get deposited at the interface continuously causing a darkening of the boundary. The paper was taken out from the suspension and dried when this darkening occurs. The interface was then digitized using a CCD camera and a frame grabber with a resolution of 260 pixels per inch (see figure 1). Since changing the room condition is found to alter the roughness of the interface even after it is dried, the humidity and temperature of the room was kept constant till the image is digitized. Figure 2 shows the behavoir of $W(l)$ against $l$ for two different values of evaporation rate. The data are averaged over 10 experiments.

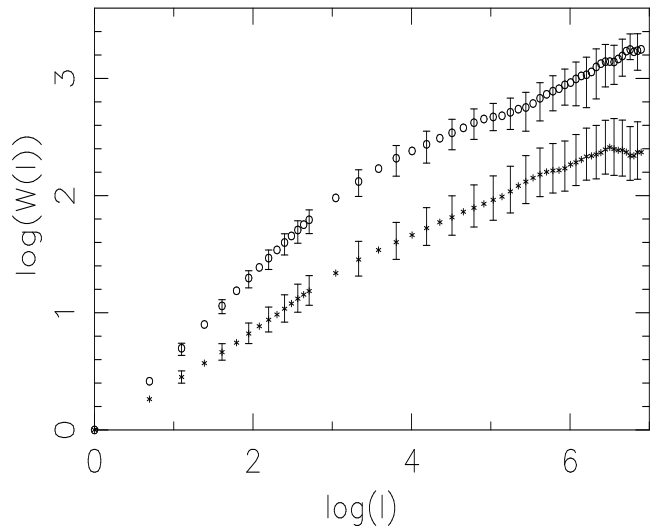

FIG. 2. The experimental values of height - height correlation function W(l) plotted against the distance of separation l after the interface stopped growing. Points marked with * fall onto a curve with exponent .45 ± .004. For lower evaporation the points (marked o ) correspond to an exponent .67 ± .004

We find the value of $\alpha$ to be $.45 \pm .004$ and $.67 \pm .004$ for higher (humidity 50.0% and temperature $21.0^\circ C$) and lower evaporation (humidity 60.0% and temperature $23.0^\circ C$) respectively. However, we do not have a quantitative measure of the evaporation. Thus $\alpha$ is very different for the two cases. A dependence of $\alpha$ on the external parameter is also exhibited by DPD models when the growth is stopped before it reaches $p_c$. There the exponent crosses over to the KPZ value. The change in $\alpha$ shown by the experiments above is different from this cross over since we see values of $\alpha$ below KPZ ($\alpha = .5$) and above DPD ($\alpha = .63$). This shows that unlike the results of reference [6] $\alpha$ is not universal and thus cannot be described by DPD model.

## III. THE MODELS

To understand the dependence of the various parameters affecting the growth of interface we present two models for studying the imbibition of a suspension into 2D random porous media. Before discussing the models used in the simulation let us try to understand the problem from a microscopic point of view. In the experiments described above paper was used as the random medium. At a microscopic level one can regard the paper as a randomly disordered medium [6] with a quenched randomness given by a fixed probability $p$ for the pores to be blocked. So, the interface growth phenomena is nothing but the propagation of fluid particles through this disordered medium. The wetting front of the fluid particles propagates due to the capillary forces. The disorder in the medium and the evaporation tries to pin this growth. It should be noted that for $p < p_c$ this pinning is possible only when the evaporation is present. Evaporation constantly decreases the number of fluid particles in the wetting front. This makes it more difficult for the fluid to overcome the obstacles. The front stops when the number of fluid particles there goes to zero. The average number of fluid particles in the front can be taken as a reduced driving force $f = (F - F_c)/F_c$. The front stops moving when $f \to 0$ at a critical height. It is true that smaller the evaporation rate, larger will be this critical height.

### A. Model I : Cellular Automaton

In this model the porous medium is considered as a square lattice with disorder being incorporated by blocking some cells randomly with a constant probability $p < p_c$ (see figure 3). The maximum capacity of each cell is fixed to $N_0$ number of fluid particles i.e. the cells are of fixed volume and the fluid is incompressible. At every time step, evaporation was explicitly modelled by the loss of certain number of particles $n$ in the transfer. At time t=0, at the bottom edge of the lattice a horizontal line of wet cells with $N_0$ particles is created. At t+1 the particles are imbibed into all unblocked cells which are nearest neighbors to the wet region. If $N$ is the number of particles in a wet cell then it transfers $N - n$ particles to all its unblocked nearest neighbours i.e. each of the neighbours get $N - n$ particles, $n$ being the loss due to evaporation. This updating is done parallely i.e. all cells which are nearest neighbour to the front are updated simultaneously. When a cell transfers to its nearest neighbours the number of particles it contains remains the same due to the source below. If a cell has more than one wet nearest neighbour it gets particles from all of them subject to a maximum number $N_0$. We also apply the rule that every cell blocked or unblocked below a new wet cell become wet as well [6]. This is to avoid the presence of overhangs and islands. We use periodic boundary condition in $x$ by identifying the cells at the edges of the lattice.

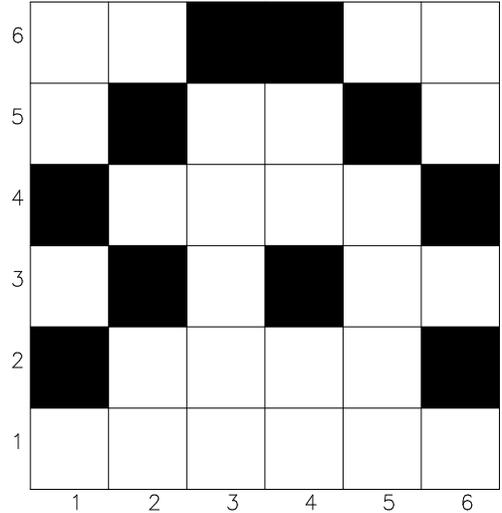

FIG. 3. Example of the multiple connectivity of the model for a 6X6 lattice. The blocked cells are shown black. At time 0 all the cells corresponding to i=1 and j=1,6 are filled with $N_0$ particles. At each time step the particles are transfered to nearest neighbours. Note that the cell (4,4) gets particles from both (4,3) and (4,5). Also when a cell ($i_s$,j) is wet all the other cells (i,j) with i $<$ $i_s$ are wet as well.

The concentration of the suspension and the pore size are incorporated in the model in the following way. For a given concentration the ratio of the number of fluid particles to the number of suspended particles in a cell is fixed. Since $N_0$ is the maximum allowed number of *fluid* particles in a cell and a cell is filled with both the *fluid* and *suspended* particles a change in concentration of the suspension effectively changes the number $N_0$. Thus a change in concentration of the suspension can be modelled by changing the maximum allowed number of fluid particles in a cell $N_0$.

In this model we can apply an external bias which could be present due to gravity or anisotropy in the medium. This bias was incorporated in the model by introducing a difference in the number of particles transfered to the vertical and horizontal neighbours of a given cell.

There are two independent parameters in the model. They are (i) the probability $p$ and (ii) the ratio $\eta = n/N_0$.

The height-height correlation function $W(l)$ in this model exhibits a power law with a nonuniversal exponent $\alpha$ which changes continuously with $\eta$ [7]. The model is different from that of reference [6], in the sense that here the effect of evaporation is incorporated in an explicit way. Unlike the directed percolation models where the interface is pinned by the connected cluster of blocks here it stops only when the wetting front runs out of fluid due to evaporation. When this happens a connected cluster of dry cells (which could be blocked or unblocked) forms at the boundary. It should be noted that the running out of fluid is qualitatively different from stopping the DPD growth before it reaches $p_c$. This is clear from the fact that for all values of evaporation the interface stops forming a connected cluster of dry cells. The average number of particles in the wetting front can be considered as the reduced driving force $f$. Thus evaporation continuously changes this driving force. Depending on the nature of the change in driving force as a function of time, the growth behaviour of the interface also changes qualitatively. The model is also different from the Eden growth [1] in the sense that all the sites in the boundary moves at the same time i.e. here the growth process by itself does not cause interface roughness.

The model can be used to study, for example, *thin layer chromatography* wherein a mixture of different chemical compounds are made to diffuse through a porous medium resulting in their separation.

### B. Model II : Random Walk

The qualitative behaviour of the interface is not very sensitive to the microscopic details of model I. This can be demonstrated by the fact that a random walk model with a finite birth - death probability captures all the features of imbibition qualitatively. The model is as follows. Consider a variable $\rho_i$. This can be thought of as the occupation number or the height of each column $i$ in a square lattice. To start with we set all $\rho_i = 0$. The cells at the tip of the columns (boundary cells) are considered as random walkers. To make exact correspondance with model-I we update all the columns simultaneously. The boundary cells either die or diffuse with fixed rates $d$ and $1 - d$ respectively.

If diffusion is chosen, then one of it's nearest neighbour is picked as the target with equal probability. If a cell up or down the boundary cell of column $i$ is chosen, then $\rho_i$ is changed by $\pm 1$. However, if the picked cell is one at left or right then two events are possible. If the neighbour is a dead one then it is regenerated with a probability $b$. In the case of alive neighbour the value of $\rho$ for both the columns are set equal to the larger one.

If death was chosen instead of diffusion then the value of $\rho_i$ corresponding to that column is kept the same. This column is then not considered for updating untill regenerated by one of its live neighbours. The growth process stops when all the boundary cells are dead.

The function $<\rho_i \rho_{i+l}>_i^{1/2}$ varies as $l^\alpha$. The exponent $\alpha$ in this model is a continuous function of the death rate $d$. We find that there is a mapping between model-I and model-II with $d \Leftrightarrow p$ and $1 - b \Leftrightarrow \eta$. The equivalent of reduced driving force $f$ here is the number of live cells at the boundary. All the qualitative features of model-I can also be obtained from model-II. In the discussion below we consider in detail model-I only. Another interesting feature of imbibition brought out by this model is it's close similarity with reaction diffusion systems which has been studied in great detail by many authors [16].

## IV. RESULTS AND DISCUSSION

In figure 4 we show the behaviour of $W(l)$ as a function of $l$ for different values of evaporation for a fixed value of $p$ and lattice size $L$. The simulations show the existence of a crossover length $l_x$ such that the height - height correlation function $W \sim l^\alpha$ for $l \ll l_x$. Whereas for $l \gg l_x$, $W$ saturates to a constant value $W_{sat}$. The exponent $\alpha$ is a function of evaporation $\eta$ while the value of $W_{sat}$ and $l_x$ depend on both the system size $L$ and evaporation $\eta$. The dependence of $\alpha$ and $W_{sat}$ on evaporation is consistent with that found in the experiment as can be seen by comparing figures 2 and 4.

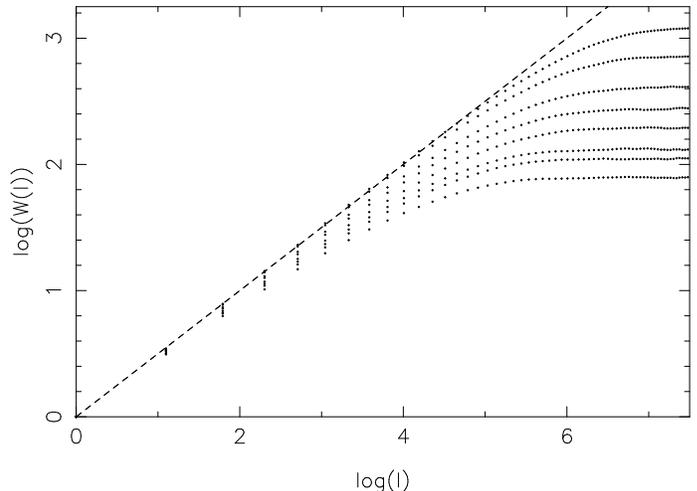

FIG. 4. The simulated height-height correlation function $W(l)$ plotted against the distance of separation $l$ after the columns stopped growing. The length is measured in units of lattice parameter. The parameters are $p = .45$, $\eta = .12$ to .148 in equal intervals of .004. The saturation value of $W(l)$ decreases monotonically with $\eta$. The simulations are done with a lattice of length 6000 applying periodic boundary condition.

The exponent $\beta$ defined through the time correlation function $C(\tau)$ as

$$C(\tau) = \langle (h(t+\tau) - h(t))^2 \rangle_t^{1/2} \sim \tau^\beta \qquad (4)$$

is shown in figure 5 for various values of $\eta$. Note that like $\alpha$, $\beta$ also exhibits a dependence on evaporation.

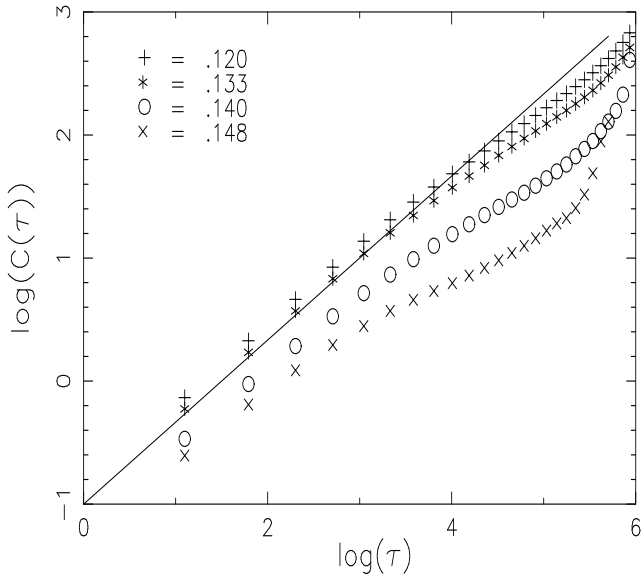

FIG. 5. The simulated time correlation function of height for values of $\eta = .120, .133, .140, .148$ and other parameters same as figure(4)

As mentioned earlier a crossover of $\alpha$ from .63 to 0.5 is seen in DPD model [5]. However, the dependence of the exponents on evaporation seen in the present model is different from this crossover. This is supported by the following arguments. (i) In both simulation and experiment we find the exponent $\alpha$ to vary from a value below 0.5 (KPZ) to one above 0.63 (DP), (ii) The dynamic exponent $\beta$ obtained from the simulation is not the same as the KPZ value in the region where $\alpha = 0.5$ (iii) Unlike the crossover region observed in DPD models here the growth is stopped only when $F = F_c$ (iv) We find $\alpha$ to vary continuously with $\eta$, i.e. we do not find any wide enough region of $\eta$ where scaling with a single value $\alpha$ can be obtained and finally (v) KPZ is expected when the interface is moving with a constant velocity, like driving DP model with $p < p_c$. But in the present model even though $p < p_c$ evaporation continuously changes the velocity. These points show that the underlying physics here is different from that of the DPD and KPZ models.

To get an insight into the dependence of $\alpha$ and $W_{sat}$ on the parameters mentioned above we plot in figure 6 the change in reduced driving force $f$ as a function of time. It is shown that for the range of $\eta$ considered $f$ behaves in the following form

$$f(t) \sim \exp[A/(t + t_c)]. \qquad (5)$$

Here $A = A(\eta)$, is a function of evaporation which sets the initial time scale. This does not affect the long time behaviour. $t_c$ is a constant for a given value of $p$. Changing $p$ does not change the behaviour of $f$ qualitatively but will change the values of $A$ and $t_c$. This nonlinear dependence of $f$ on time explains why the interface behaviour is different from that of DPD model in ref [6].

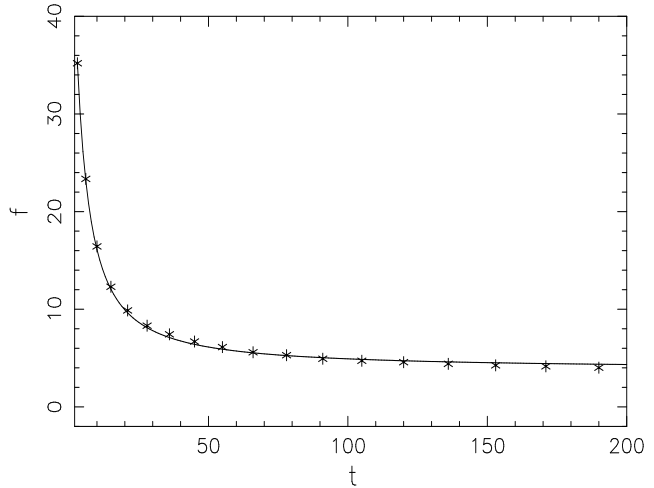

FIG. 6. The nature of the reduced driving force $f$ for $\eta = .136$ and $p = .45$. The continuous line is the fit to equation (5) with $t_c = 10.0$ and $A = 29.0$

The driving force $f$ saturates to a value $f_s$ at $t \sim 2t_c$. The value of $f_s$ as a function of $\eta$ is given in figure 7 at $t = 2t_c$. We would like to emphasize the point that changing $\eta$ do not change $t_c$ but merely changes the value of $f_c$ at which the system is driven and $A$ which sets the transient time scale. We find that there are two critical values of $\eta$, $\eta_1$ and $\eta_2$ at which the slope of the $f_s(\eta)$ curve changes drastically. The interface behaves differently in this three regions.

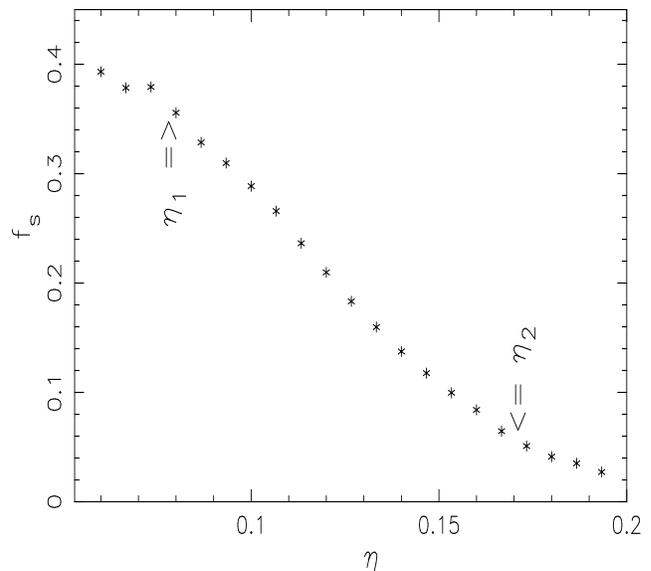

FIG. 7. The reduced driving force $f_s = f(2t_c)$ showing two critical values of $\eta$ for $p = .45$

It is easy to understand the reason for the existence of two critical values of $\eta$. In the model the particles have more than one path to reach a particular cell. The effective number of paths available to reach a cell decreases with evaporation since increased evaporation suppresses the longer paths. For $\eta < \eta_1$ the number of fluid particles that a cell loses through evaporation is more than compensated by the inflow because of the many paths available. Hence this regime becomes super diffusive. We see from figure 8 that the value of $\alpha$ in this region to be greater than .5. As the number of paths become less the particles get stuck at obstacles for a longer time inducing a transition into a normal diffusive regime with exponents close to .5. In this region the reduced driving force $f$ saturates to a non-zero value ($f_s \sim 0^+$) at $t \sim 2t_c$. This driving of the interface close to $f \sim 0^+$ gives rise to some interesting behaviour. We will discuss this region in detail separately. On further increase of evaporation the driving force $f$ goes to zero at $t = 2t_c$. The system in this region becomes subdiffusive with exponents falling much below .5.

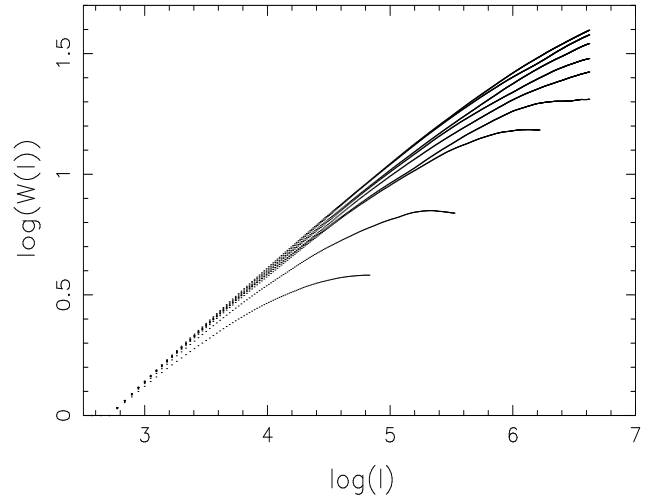

FIG. 9. The simulated height-height correlation function $W(l)$ plotted against the distance of separation $l$ after the columns stopped growing for system sizes $L = 125$, 250, 500, 750, 1000, 1250, 1500, 1750, 2000, $p = .2$, $\eta = .293$

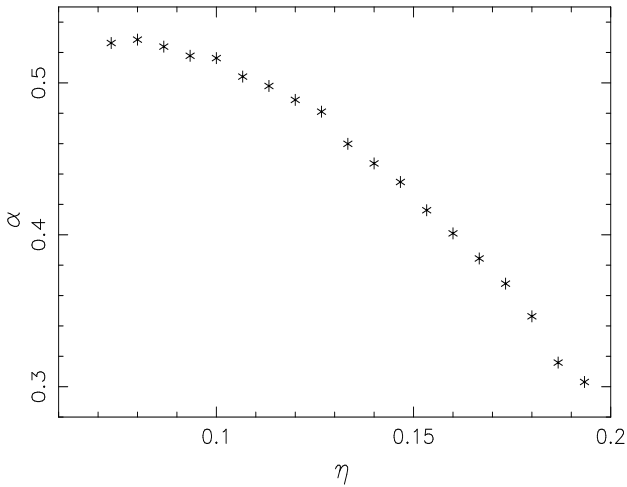

FIG. 8. The dependence of the simulated exponent $\alpha$ on $\eta$ at $t = 4t_c$ for $p = .45$

### A. Scale invariant region ($\eta_1 < \eta < \eta_2$)

This is the most interesting region. Here $W(l)$ shows a dependence on the system size $L$. This is depicted in figure 9. We find that the cutoff width $W_{sat}$ depends on the system size as

$$W_{sat} \sim L^\gamma, \qquad (6)$$

with the width function $W(l, L)$ satisfying a finite size scaling form

$$W(l, L) = L^\gamma g(l/L^\nu). \qquad (7)$$

Where $\gamma = .34$ and $\nu = .68$ for $p = .2$ and $\eta = .293$ The data collapse obtained with this scaling is shown in figure 10. As implied by equations (1) and (7) we find the exponents $\gamma$, $\nu$ and $\alpha$ satisfies the scaling relation [17]

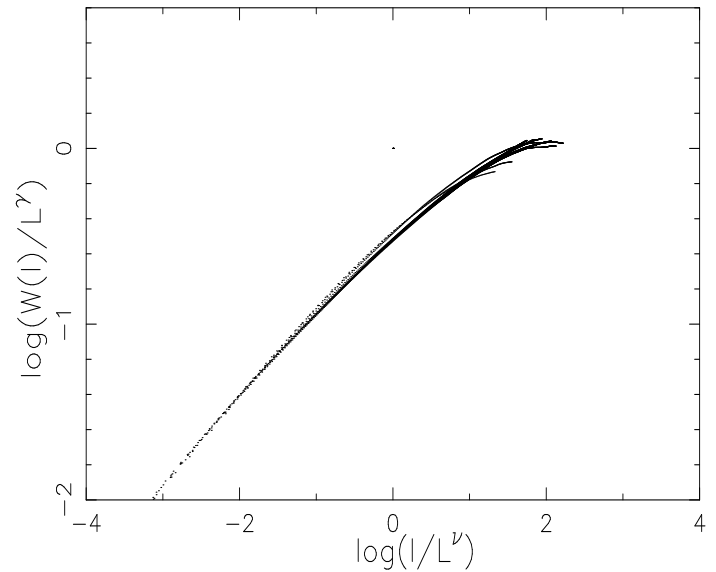

FIG. 10. The data collapse with the finite size scaling (see equation (7)) for $\gamma = .34$ and $\nu = .68$. The other parameters are same as in figure 9

$$\frac{\gamma}{\nu} = \alpha, \qquad (8)$$

since for the given value of $p$ and $\eta$ $\alpha = .52$. This system size scaling establishes that there is no single characteristic length scale in the problem. From our simulations for the same value of $p$ we find another relation

$$\sum_{l=1}^{L} W(l, L) = L^\mu \qquad (9)$$

However the scaling relation $\gamma + \nu = \mu$ obtained from equations (7) and (9) is not obeyed here pointing to multifractal behaviour. This implies that the dominant nature of the function $W(l, L)$ is the power law relation given by equation (1) [17].

As mentioned earlier the interface stops moving when all the boundary cells dry up. Then a connected cluster of dry cells is formed at the boundary. We show in figure 11 that the length distribution of this connected clusters approches a power law distribution,

$$P(\xi) \sim \xi^{-\kappa} \qquad (10)$$

as $t \to \infty$. This power law dependence is typical of scale invariant systems.

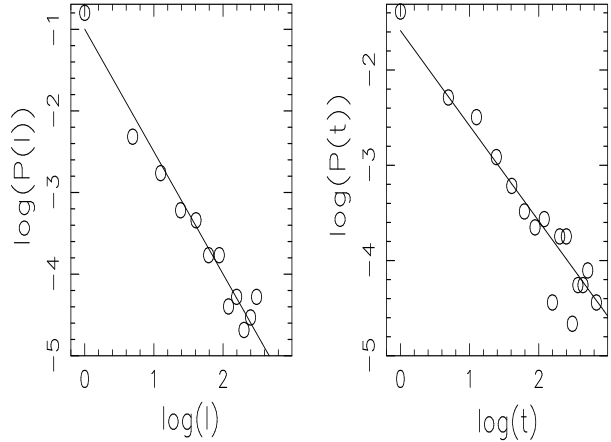

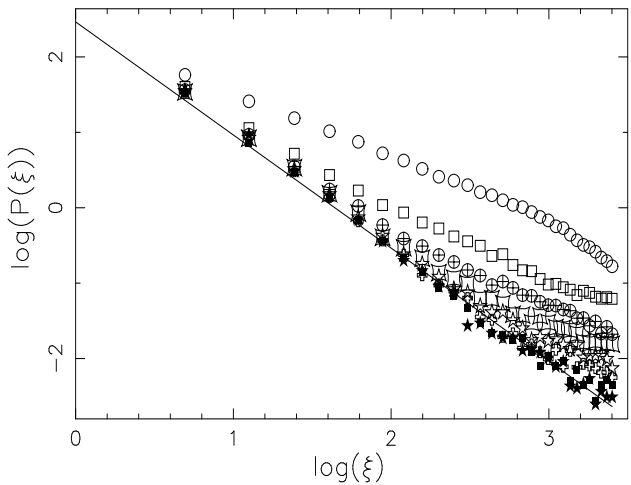

FIG. 11. Distribution of the length of connected cluster of dry cells calculated at $t = t_c$ to $t = 8t_c$. Note that the distribution approaches to a power law given by $P(\xi) \sim \xi^{-1.5}$.

### B. self-organised criticality (SOC)

To establish the SOC nature of the interface in the above models it is necessary to obtain the length and time distribution of avalanches [18]. We get this by the following method. When the interface stops moving a connected cluster of dry cells is formed at the boundary. We then fill one of these dry cells with $N_0$ number of particles. This results in further movement of the interface. The spatial and temporal distribution of these avalanches is shown in figure 12. These distributions are found to obey the power laws

$$P(l) \sim l^{-\kappa} \qquad (11)$$

$$P(t) \sim t^{-\delta}. \qquad (12)$$

FIG. 12. Spatial and temporal distribution functions of the avalanches shown for the parameters same as in figure 6

The exponent $\delta$ has a universal value equal to 1.0 for $\eta_1 < \eta < \eta_2$. It has been argued recently by Boer et al. that the mere existence of power law correlations is not sufficient to establish SOC [18]. They take the non-normalisability of the temporal distribution function to be the criterion of SOC. The corresponding argument here is the following. For an infinite lattice the finiteness of the integral $\int_1^\infty P(t)dt$ means that points infinitely far from the origin of avalanche do not participate in the avalanche process. If the interface is *critical* it is robust against any disturbance. *i.e.* an avalanche should leave the interface unchanged. This is possible only if all the points on the interface has a nonzero probability of participating in the avalanche indicating $P(t)$ as $t \to \infty$ to be significant. This results in the divergence of the inegral. The value of $\delta = 1.0$ in our simulation satisfies this criterion establishing the SOC of the interface.

There exists a simple relation between the exponents $\kappa$ and $\alpha$ given by $\kappa = 1.0 + \alpha$. This can be seen from the following. The total area of an avalanche of size $l$ is $\int_0^l l^\alpha dl \sim l^{1+\alpha}$. Since there is a uniform probability of blocking the probability of the avalanche of size $l$ is $l^{-(1+\alpha)}$. The same argument applies for the distribution $P(\xi)$ shown in figure (11).

To understand this critical behaviour of the interface we look at the nature of the driving force. Remember, (see figure 6) the time dependence of the driving force is of the form given in equation(5). This means that the interface is driven by a force $F \sim F_c$ for a long time. It should be noted that the tuning of the driving force to $F_c$ is done by the system internally. This is a characteristic feature of self organising systems, wherein the system is driven in a region close to criticality by itself [19]. This shows that in this region of $\eta$ the interface exhibits self organisation.

The behaviour of correlation function $W(l)$ on evapo-ration is similar to the dependence of distribution functions on nonconservation in cellular automaton in earthquake models [14]. In fact the dependence of the exponent $\alpha$ on evaporation is strikingly similar to that of the exponent B on nonconservation parameter in reference [14]. This indicates that imbibition is better described by cellular automaton models than directed percolation. The main points which distinguishes the present model from that of DP are the following. (i) Unlike the present model DP model has universal exponents, (ii) the DP model is driven to $p_c$ in a linear fashion whereas the present model has a nonlinear driving force leading to a terminal driving of the interface and most importantly (ii) we find the avalanches to show power law distribution while the avalanches in the DP model shows no interesting scaling behaviour and has an exponential distribution [20]. This power law distribution shows that evaporation does not set any length scale contrarary to the DP model wherein there is a length scale set by the value of $\Delta p$.

### C. Multifractality

To check for multiscaling of the interface we calculate from our simulation the $m_{th}$ order height-height correlation function defined by [21]

$$W_m(l) = \langle |h(x+l) - h(x)|^m \rangle \sim l^{m\alpha_m}. \tag{13}$$

The value of the exponent $\alpha_m$ for various values of $m$ is shown in figure 13. As is well known for multifractal surfaces we find $\alpha_m$ to change continuously with $m$.

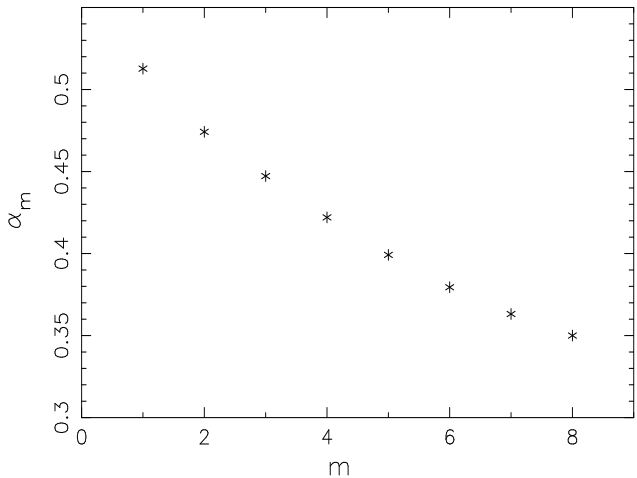

FIG. 13. The exponent $\alpha_m$ plotted against $m$. Parameters are $p = .45$ and $\eta = .136$

We also carry out a multifractal analysis using the height distribution $N(j)$ [22,23]. Here $j$ is the deviation of height of the columns from the mean h. We define a normalised probability distribution

$$P(j) = \frac{N(j)}{\sum_{j=-M}^{M} N(j)} \tag{14}$$

Then the singularity spectrum $f(\theta)$ is defined as

$$f(\theta) = \frac{-1}{\ln(2M+1)} \sum_{j=-M}^{M} W(j) \ln W(j) \tag{15}$$

where,

$$W(j) = \frac{P(j)^m}{\sum_{j=-M}^{M} P(j)^m}$$

and

$$\theta = \frac{-1}{\ln(2M+1)} \sum_{j=-M}^{M} \ln W(j)$$

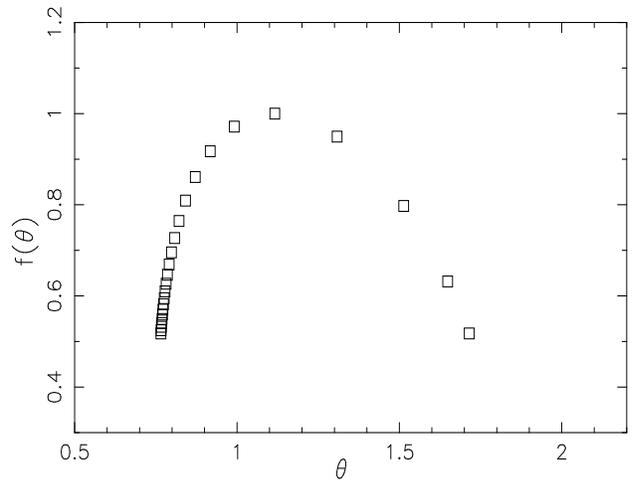

FIG. 14. $f(\theta)$ versus $\theta$ spectrum for the same parameters as figure 13

In figure(14) we depict the smooth dependence of $f(\theta)$ on $\theta$. Remember for ordinary fractals this curve degenerates into a single point.

It is to be noted that the results are presented for a particular choice of parameters in the model. A change in the parameters does not alter the gross features.

### V. CONCLUSION

The main points of this paper are the following.
(i) We have shown from experiments conducted at two different conditions that the exponent $\alpha$ in imbibition is not universal.

(ii) A model for imbibition (model I) which includes the effect of evaporation through loss of particles on transfer is presented.

(iii) The model shows that the static height - height correlation exponent $\alpha$ depends on the evaporation which is consistent with our experiments.

(iv) A random walk model (model II) which captures all the qualitative features of imbibition is presented to show the independence of the general results on microscopic details.

(v) The interface is shown to exhibit self-organised criticality for a region of evaporation. The dependence of the exponents on evaporation is simiar to that of exponents in nonconserved cellular automaton indicating that imbibition falls into this class.

(vi) It is shown that the driving force tunes itself close to the critical value for all values of evaporation $\eta_1 < \eta < \eta_2$. This slow driving of the system close to pinning can be understood as the reason for the existence of self-organisation [19].

(vii) The interface also shows multifractality.

To conclude, imbibition experiments show that the roughness exponents depend on evaporation. A first principles cellular automaton model for imbibition gives results in agreement with experiments. Also the model exhibits properties much richer than the already known models of imbibition.

## ACKNOWLEDGMENTS

Our thanks are due to Jayadev Rajagopal and S. Chanthrasekaran for their help with the image acquisition and digitization.

* Present address: Institute of Mathematical Sciences, Madras 600 113, India. E-mail: sunil@imsc.ernet.in
† E-mail: deb@rri.ernet.in

[1] T.H. Healy and Y.-C. Zang, Physics Reports., **254**, 215 (1995)
[2] T. Vicsek, *Fractal Growth Phenomena* (World Scientific, Singapore, 1992), 2nd ed., Pt IV; *Solids Far From Equlibrium: Growth, Morphology and defects*, edited by C.Godreche (Cambridge University Press, Cambridge, England, (1991)
[3] J.W. Evans, Rev. Mod. Phys., **65**, 1281 (1993)
[4] M.A. Rubio, C.A. Edwards, A. Dougherty, and J.P. Gollub, Phys. Rev. Lett., **63**, 1685 (1989)
[5] A.L. Barabási, S.V. Buldyrev, S. Havlin, G. Huber, H.E. Stanley and T. Vicsek, in *Surface Disordering: Growth, Roughening and Phase Transitions* edited by R. Jullien, J. Kertsez, P. Meakin and D.E. Wolf, Proceedings of the Les Houches Workshop, 1992 (Nova Science, New York, 1992), L.H. Tang and H. Leschhorn, Phys. Rev. A., **45**, R8309 (1992), S.V. Buldyrev, A.-L. Barabási, F. Caserta, S. Havlin, H.E. Stanley and T. Vicsek Phys. Rev. A., **45**, R8313 (1992)
[6] L.A.N. Amaral, A.L. Barabási, S.V. Buldyrev, S. Havlin and H.E. Stanley, Phys. Rev. Lett., **72**, 641 (1994), L.A.N. Amaral, A.L. Barabási, S.V. Buldyrev, S.T. Harrington, S. Havlin, R. Sadr-Lahijany, and H.E. Stanley. Phys. Rev. E **51**, 4655 (1995)
[7] P.B. Sunil Kumar and Debnarayan Jana, *Imbibition: Experiment and Simulation*, To be published in proceedings of "Dynamics of Complex system", Satelite meeting to STATPHYS-19 (1995).
[8] *Random Fluctuations and pattern growth: Experiments and Models*, edited by H.E. Stanley and N. Ostrowsky, NATO ASI Series E:Vol 157, (Kluwer Academic Publishers, 1988)
[9] M.C. Cross and P.C. Hohenberg, Rev. Mod.Phys., **65**, 851 (1993)
[10] M. Kardar, G. Parisi and Y.-C. Zhang, Phys.Rev.Lett., **56**, 889 (1986)
[11] L.A.N. Amaral, A.L. Barabási and H.E. Stanley, Phys. Rev. Lett., **73**, 62 (1994),L.A.N. Amaral, A.L. Barabási H.A. Makse and H.E. Stanley, *To appear in* Phys. Rev. E
[12] P. Bak, C. Tang and K. Weisenfield, Phys. Rev. Lett. **59**, 381 (1987), Phys. Rev. A, **38**, 364 (1988)
[13] G. Grienstein, D.-H. Lee and S. Sachdev, Phys. Rev. Lett. **64**, 1927 (1990)
[14] Z. Olami, H.J.S. Feder and K. Christensen, Phys. Rev. Lett., **68**, 1244 (1992). K. Christensen and Z. Olami, Phys. Rev. A, **46**, 1829 (1992)
[15] A.A. Middleton and Chao Tang. Phys. Rev. Lett. **74**, 742 (1995)
[16] V. Kuzovkov and E. Kotomin, Rep. Prog. Phys., **51**, 1479 (1988)
[17] L.P. Kadanoff, S.R. Nagel, L. Wu and S. Zhou, Phys. Rev. A., **39**, 6524 (1989)
[18] J. de Boer, A.D. Jackson and Tilo Wettig, Phys. Rev. E, **51**, 1059 (1995)
[19] D. Sornette, A. Johansen and I. Dornic, J. Phys. I France **5**, 325 (1995)
[20] Z. Olami, I. Procaccia and R. Zeitak, Phys. Rev. E, **49**, 1232 (1994)
[21] A.-L. Barabási and T. Vicsek, Phys. Rev. A, **44**, 2730 (1991), A.-L. Barabási, R. Bourbonnais, M. Jensen, J. Kertész, T. Vicsek and Y.-C. Zhang, Phys. Rev. A, **45**, R6951 (1992)
[22] T.C. Halsey, M.H. Jensen, L.P. Kadanoff, I. Procaccia, and B.I. Shraiman, Phy. Rev. A. **33**, 1141 (1986)
[23] P.K. Thakur and C. Basu, Physica A, **216**, 45 (1995)